\documentclass[%
 aip,
 amsmath,amssymb,
 reprint,%
]{revtex4-1}

\usepackage{qcircuit}
\usepackage{graphicx}% Include figure files
\usepackage{dcolumn}% Align table columns on decimal point
\usepackage{bm}% bold math
\usepackage[utf8]{inputenc}
\usepackage[T1]{fontenc}
\usepackage{mathptmx}
\usepackage{xcolor}
\usepackage{tikz}
\usepackage{caption}
\usepackage{wrapfig}
\newcommand{\Tr}{\text{Tr}}
\newcommand{\A}{\mathcal{A}}
\newcommand{\id}{\mathbf{1}}
\newcommand{\ad}{a^\dagger}
\renewcommand{\eqref}[1]{eq.~(\ref{#1})}
\definecolor{ao(english)}{rgb}{0.0, 0.5, 0.0}
\begin{document}

\preprint{AIP/123-QED}

\title[HF-QC]{Limitations of Hartree-Fock with quantum resources}
% Force line breaks with \\

\author{Sahil Gulania}
 \email{gulania@usc.edu}
 \affiliation{Department of Chemistry, University of Southern California, Los Angeles, CA, 90089}%Lines break automatically or can be forced with \\
\author{James Daniel Whitfield}%
 \email{james.d.whitfield@dartmouth.edu}
 \affiliation{Department of Physics and Astronomy, Dartmouth College, Hanover, NH 03755}

\date{\today}% It is always \today, today,

\begin{abstract}
The Hartree-Fock problem provides the conceptual and mathematical underpinning of a large portion of quantum chemistry.  As efforts in quantum technology aim to enhance computational chemistry algorithms, the fundamental Hartree-Fock problem is a natural target.  While quantum computers and quantum simulation offer many prospects for the future of modern chemistry, the Hartree-Fock problem is not a likely candidate.  We highlight this fact from a number of perspectives including computational complexity, practical examples, and the full characterization of the energy landscapes for simple systems. 
\end{abstract}
\maketitle

The present study aims to highlight the difficulty of conducting optimization problems in the context of electronic structure, with and without quantum resources. Specifically, we will focus on  Hartree-Fock, an optimization problem using the mean-field approximation.
The Hartree-Fock problem \cite{Hartree28,Fock30} provides the mathematical setting for molecular orbitals widely used in chemistry and beyond. 
The ubiquitous self-consistent field (SCF) methodology used to solve Hartree-Fock is also applied to most implementations of density functional theory based on Kohn-Sham theory \cite{Kohn65}. 
While the solution to an instance of the Hartree-Fock problem is often insufficient for many applications, it
often serves as the  
reference state for post-Hartree-Fock methods. Widely used post-Hartree-Fock methods include coupled cluster ansatz~\cite{coester1958bound,coester1960short,vcivzek1966correlation,bartlett1981many,paldus1999critical,shavitt2009many},  Møller–Plesset perturbation theory~\cite{kutzelnigg2009many,lindgren1982springer,binkley1975moller,cremer2011moller}, equation of motion~\cite{rowe1968equations,emrich1981extension,geertsen1989equation,stanton1993equation,levchenko2004equation,sinha1986note,krylov2008equation}, multi-reference configuration interaction~\cite{werner1988efficient,knowles1989determinant,knowles1992internally} and many more.  Here, we forgo improving the Hartree-Fock ansatz and instead ask how difficult it is to find the true Hartree-Fock global minimum and its importance.

Many instances of the Hartree-Fock problem can be solved quickly using heuristic approaches.  
In practice, conventional algorithms for Hartree-Fock scales cubically with the number of basis functions.
However, this cost only reflects the correct scaling if the number of iterations is bounded by a constant or if local minimums are acceptable in place of a global solution.  
Linear scaling methods~\cite{Goedecker99,Zalesny11} avoid the diagonalization of hessian
entirely and rely on localized properties of the system.
This assumption may not be true generally for every case. 
Iterative procedures, regardless of cost per step, are prone to convergence issues.
This article highlights the above mentioned properties in the context of standard numerical and hybrid-quantum approaches to the Hartree-Fock problem.  
 % Regardless, these methods have yet to mature to the point where they can supplant ordinary implementations. 
 
Typical approaches to improve SCF convergence uses 
direct inversion of the iterative subspace~\cite{Pulay80}(DIIS),
level shifting~\cite{Saunders73}, quadratically convergent 
Newton-Raphson techniques~\cite{Bacskay81}, or varying 
fractional occupation numbers~\cite{Rabuck99}, among many other 
approaches. 
The success of any above mentioned methods varies at each instance and depends on the initial parameters chosen, e.g.~the size of the iterative subspace, and often work well in combination leading to attempts to build black-box SCF procedures~\cite{Thogersen04,Kudin02}.   

Unfortunately, these methods cannot work in all cases without violating fundamental assumptions in the theory of computation. In previous work, arbitrary spin-glasses have been mapped to instances of the  Hartree-Fock problem \cite{Whitfield14b}.  In addition to other works \cite{Schuch09,Whitfield14a}, this shows that the  Hartree-Fock problem is difficult in the worst case setting.  The complexity class of non-deterministic polynomial time (NP) problems are the set of problems \footnote{In this article, we use the term \emph{problem} to refer to a specific collection of problem instances} that can be solved efficiently with a hint \cite{Sipser97}.  It is possible that every problem that can be solved efficiently with a hint, can also be efficiently solved without the hint. However, from all signs of practical experience suggest that problems in the NP class can do not admit efficient black-box solutions.  Thus, due to the NP-completeness of the Hartree-Fock problem, it is unlikely that any classical algorithm can solve all instances in a time proportional a polynomial of the input size.  Note that because it in the NP complexity class, the Hartree-Fock problem can be solved efficiently in polynomial time with a sufficiently strong hint e.g. as gleamed from experience or luck. It has been known since the late 1990's that  quantum computers can promise no more than a quadratic speed up over their classical counterparts on such NP-complete problem \cite{Bennett97}.  %This article will not uncover that quadratic speed up since its usage is likely i

It has long been known that any classical algorithm can be simulated using quantum hardware.  With the the heavy reliance of the variational quantum eigensolvers on classical optimization routines, it is marginal how the quantum approach differs from the conventional approaches.  In the recent VQE study the optimization strategy used was an augmented Hessian approach \cite{Sun16} rather than the typical DIIS \cite{Pulay80}.  However, both of these optimization strategies can be employed by conventional computers to optimize the orbitals.

In this brief communication, we consider how the use of quantum hardware enhances the ability of chemists to solve instances of the Hartree-Fock problem.  Quantum hardware has many prospects for applications to physical and chemical simulations \cite{Lloyd96}, however the Hartree-Fock problem is not likely to admit drastic advances using quantum computers.  

The present article is inspired by a recent study of the Google group and collaborators \cite{Google20} using a variational quantum-classical hybrid formulation of the Hartree-Fock problem.  There the authors point out the primary purpose of considering the Hartree-Fock problem was to benchmark their quantum device.  Here, we highlight obstructions to its use as a general purpose replacement for standard SCF solvers.  We do so by first introducing the Hartree-Fock problem and characterizations of the solution landscape for an instance of the Hartree-Fock problem.  We then give examples that are (1) simple, (2) small, and (3) well-motivated instances of Hartree-Fock that present convergence problems for black-box approaches.

\section{\label{sec:ansatz}Hartree-Fock theory}
The Hartree-Fock ansatz is important for conventional quantum chemistry and has been thoroughly developed over the past 100 years \cite{Canes03}.  The Hartree-Fock problem can be stated succinctly as: minimize the electronic energy in the space of single Fock states.
\begin{equation}
	E_{HF}=\min_{\Psi\in \mathcal{F}_1}\langle \Psi|\left( \sum_{ij}h_{ij}a^\dag_ia_j+\sum_{ijkl}h_{ijkl} a^\dag_ia^\dag_ja_ka_l\right) |\Psi\rangle
	\label{eq:HF1}
\end{equation}

The space $\mathcal{F}_1$ is the set of all rank 1 (versus rank $M \choose N$ in the general case) $N$-electron Fock states. Here and throughout, $N$ is the number of electrons and $M$ is the number of spatial basis functions.  Each single Fock state is of the form: 
$\Psi=b_1^\dagger b_2^\dagger \cdots b_N^\dagger|{\Omega}\rangle$ where we have $[b_i,b_j^\dag]_+=b_{i}b_j^\dag+b_j^\dag b_i =\delta_{ij} $ and $[b_j,b_k]_+=0$.

After the restriction to this ansatz, the variational space is characterized by rotations from an initial set of orbitals (e.g. the atomic orbitals).  If the initial set corresponds to operators $\{a_j : [a_i,a_j^\dag]_+=\int dx\;\phi_i(x)\phi_j^*(x)=S_{ij}, [a_i,a_j]_+=0\}$, then
\begin{equation}
	b_i^\dag=\sum_j W_{ji}\ad_j.
	\label{eq:MO}
\end{equation}
Here we require that $WW^\dag=\id$ only when $S=\id$. For applications to molecular physics, single-electron spin-orbitals $\phi_i(x)$ corresponding to the fermionic operators $\ad_j$ and $a_j$ are of Gaussian form e.g. STO-3G \cite{sto3g}.

\subsection{Rotation of charge density matrix}
 
In practice, most algorithms utilize the SCF method to solve 
\eqref{eq:HF1} using a effective potential term that takes into account the averaged two-body interaction (mean-field).  In this approach, the $N$-body problem is reduced to a non-linear single particle 
problem. At each iteration of the simplest implementation,  the Fock 
matrix, $F$, is formed as a function of a previous bond density matrix ($D_{prev}$)
\begin{equation}
    D_{prev}=[\langle a_{j\uparrow}^\dag a_{k\uparrow}+ a_{j\downarrow}^\dag a_{k\downarrow}\rangle_\psi]_{jk}^M= C_{prev}\eta C^\dag_{prev}
\end{equation}
with $\eta$ as the orbital occupancies written as a diagonal matrix.
The new transformation matrix, $C$, is determined using the gradient of \eqref{eq:HF1} with respect to bond density matrix $D$. The new coefficient matrix is used to form a new bond density 
matrix.  At each iteration, the (real-valued) bond density matrix satisfies the following three properties:
\begin{eqnarray}
D&=&D^T\\
\Tr(DS)&=&N/2\\
D&=&DSD
\end{eqnarray}
We convert to an orthogonal basis using e.g. canonical orthogonalization $X_{canonical}=U_S^\dag \sqrt{s}$ with $S= U_SsU_S^\dag$. converts from the non-orthogonal basis to an orthogonal 
one.
e.g.  
canoncially with $X_{canonical}=\sqrt{S}$, $P=X^\dag DX$.

The three 
properties of the bond-density matrix below imply, in an orthogonal basis, that $P$ is a rank $N/2$ projector with trace $N/2$.  We use $D$ to denote an arbitrary bond density matrix and $P$ as a bond density matrix in an orthogonal basis.

Next, we consider transformation between bond density matrices.  A convenient parameterization of the set of bond density matrices that avoids redundancy is
\begin{equation}
P(A)=e^{-A_{block}}P_0e^{A_{block}}
\label{eq:PX}
\end{equation}
with $A_{block}=P_0A(\id-P_0)+(\id-P_0)AP_0$ for arbitrary skew-symmetric $A=-A^T$.

Once an instance of a Hartree-Fock problem has been specified, the objective for the optimization problem is given by the energy matrix functional
\begin{equation}
E[P]=2 \Tr[h P] + \Tr[ G\; P].
\label{eq:HFF}
\end{equation}
Here the mean-field, $G$, is a function of the bond density matrix:
\begin{equation}
G=G[P]_{\mu\nu}=\sum^M_{\kappa,\lambda} h^\A_{\kappa\mu\nu\lambda} P_{\kappa\lambda}
\end{equation}
with the antisymmetrized two-electron integral over spatial degrees of freedom defined by $h^\A_{\mu\kappa\lambda\nu}= h_{\mu\kappa\lambda\nu}-h_{\mu\kappa\nu\lambda}$.  Note that expression \eqref{eq:HFF} follows directly from evaluating the energy's expectation value in \eqref{eq:HF1} using e.g. Slater-Condon rules \cite{Szabo96}. The matrix derivative of \eqref{eq:HFF} gives the Fock operator
\begin{equation}
    F_{ij}=2(h_{ij}+G_{ij}).
\end{equation}

By rotating the charge density matrix with all possible rotations, we are able to do brute force exploration of whole space for some small examples.  Below we expand about $P_0=P_{core}$ where the $N$ lowest eigenmodes of $H_{core}=h_{ij}a_{i}^\dag a_j$ are occupied.
Before turning to examples, we will introduce the quantum ansatz for the Hartree-Fock method.

\section{Quantum circuit ansatz for Hartree-Fock}
The quantum circuit for creating the Hartree-Fock ansatz state \cite{Kivlichan17} was applied in the context of VQE \cite{Google20}.
The description of the ansatz circuit will be aided through the use of QR decomposition \cite{Horn05}.

\begin{figure}[t!]
    
    \begin{center}
    \includegraphics[scale=0.6]{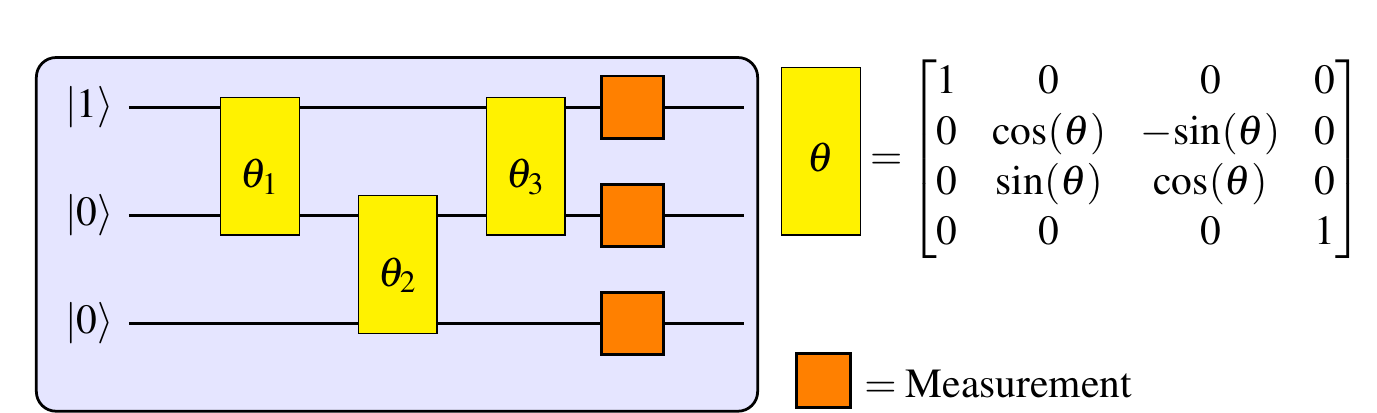}
    \end{center}
    \caption{Quantum circuit for the two fermion and three molecular orbital (for example H$_{3}^{+}$ in STO-3G basis).
    \label{fig:circuit}}       
\end{figure}

The Givens rotations provide a useful canonical characterization of an arbitrary orthogonal matrix, $W$.  The QR decomposition of a real $n\times n$ orthogonal matrix $W$ can be done using $T=n(n-1)/2$ Givens rotations such that \begin{equation}
        W=G_1 G_2 G_4... G_{T} D
\end{equation} 
When $W$ has determinant of one, $D$ is just the identity matrix.  Each Givens rotation, $G_i$, is of the form $G_i=g(a,b,\theta)$ with $g_{kk}=1$ unless $k$ is either $a$ or $b$ when instead $g_{kk}=\cos(\theta)$. All off-diagonal elements are zero except $g_{ab}=-g_{ba}=-\sin(\theta)$.

Applications of the Givens decomposition to fermionic orbital rotations has been worked out elsewhere \cite{Kivlichan17,Jiang18} resulting in a quantum circuit that is able to prepare arbitrary Slater determinants following the parameters of the QR decomposition.  By ordering the QR decomposition appropriately, a fermionic swap network can be used to rotate each pair of orbitals using the appropriate Givens rotation parameters.  This results in an efficient state preparation circuit of the form depicted in fig.~\ref{fig:circuit}.  The full compilation down to gates including hardware optimization is given elsewhere \cite{Google20,hfvqe}. 

Our characterization of the fermionic space in  \eqref{eq:PX} gives us a set of parameters, $\Theta$ that also characterizes the mixing between pairs of orbitals. The resulting orthogonal transformation $W(\Theta)$ is then given to the QR decomposition and forwarded to the quantum circuit construction.

\section{Calculations and Results}
All calculations of the molecular system are done in the STO-3G basis \cite{sto3g}. Energies are reported in Hartrees, angles of rotation in radians, and bond lengths in Angstroms. 

The Hartree-Fock energy surfaces (HES) were computed using PySCF \cite{pyscf20,pyscf17}. In this paper we only consider Restricted Hartree-Fock (RHF) solutions where the alpha and beta spatial orbitals are restricted to be identical. The quantum optimization routines were that of OpenFermion-Cirq and we only modify the initial state routines and the input molecular data \cite{openfermion}. The data that support the findings of this study are available from the  authors upon reasonable request.

\subsection{Landscape analysis}

We consider H$_2$, H$_3^+$ as minimal basis model systems whose Hartree-Fock instances we can completely characterize. We begin with the H$_2$ example.

\begin{figure}[h!]
    \centering
    \includegraphics[scale=0.6]{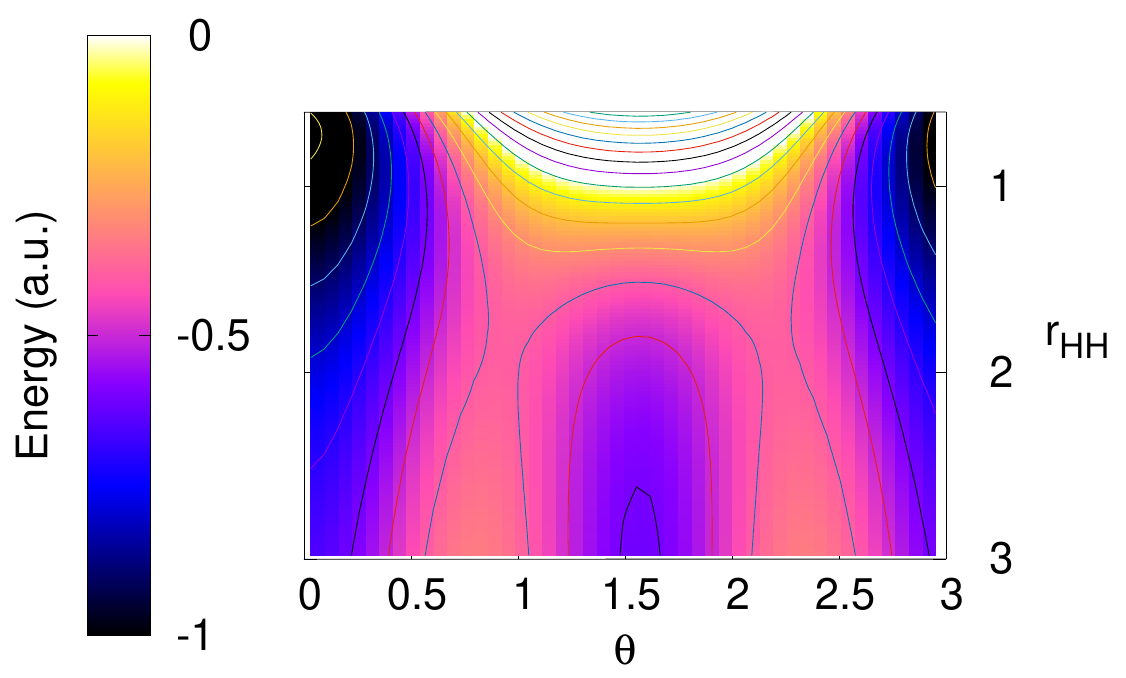}
    \caption{H$_{2}$ HES$(\theta$; r$_{\text{HH}}$). Each fixed value of the nuclear separation, r$_{\text{HH}}$, generates a Hartree-Fock instance characterized by a single the orbital rotation parameter, $\theta$. Notice the appearance of a second HES minimum at a higher energy around $r\approx 1.2$ \AA. }
    \label{fig:h2_HES_rot}
\end{figure}

When considering H$_2$ in the minimal basis with there is only a single orbital mixing parameter.  In fig.~\ref{fig:h2_HES_rot}, we have plotted the 1D HES surface as a function of bond length for H$_{2}$. The number of  minimums in
HES$(\theta$) changes with bond length. Before a bond length of approximately $1.2$ \AA, there is only a single minimum. But at larger bond lengths an additional minimum begins to appear.

\begin{figure}[th!]
    \centering
    \includegraphics[scale=0.26]{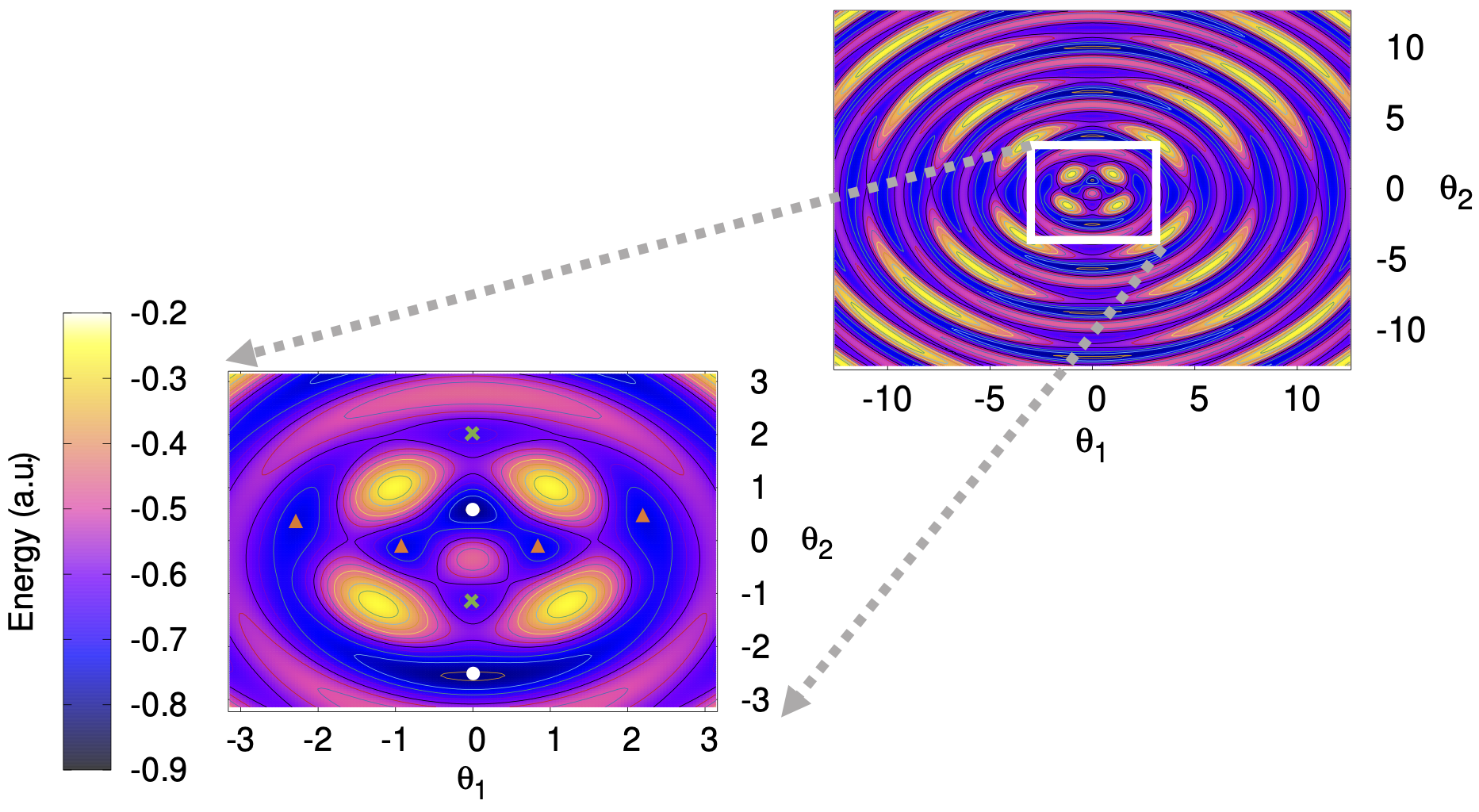}
    \caption{H$_{3}^{+}$ HES$ (\theta_{1},\theta_{2}$), showing three different minimums (global minimum - $\circ$, second minimum - ${\color{orange} \blacktriangle}$ and third minimum - ${\color{ao(english)} \times} $ ) at r$_{\text{HH}}=2.5$ \AA.  Here the surface is expanded about $P_0=P_{core}$.}
    \label{fig:h3_HES_rot}
\end{figure}

\begin{figure}[th!]
    \centering
    \includegraphics[scale=0.26]{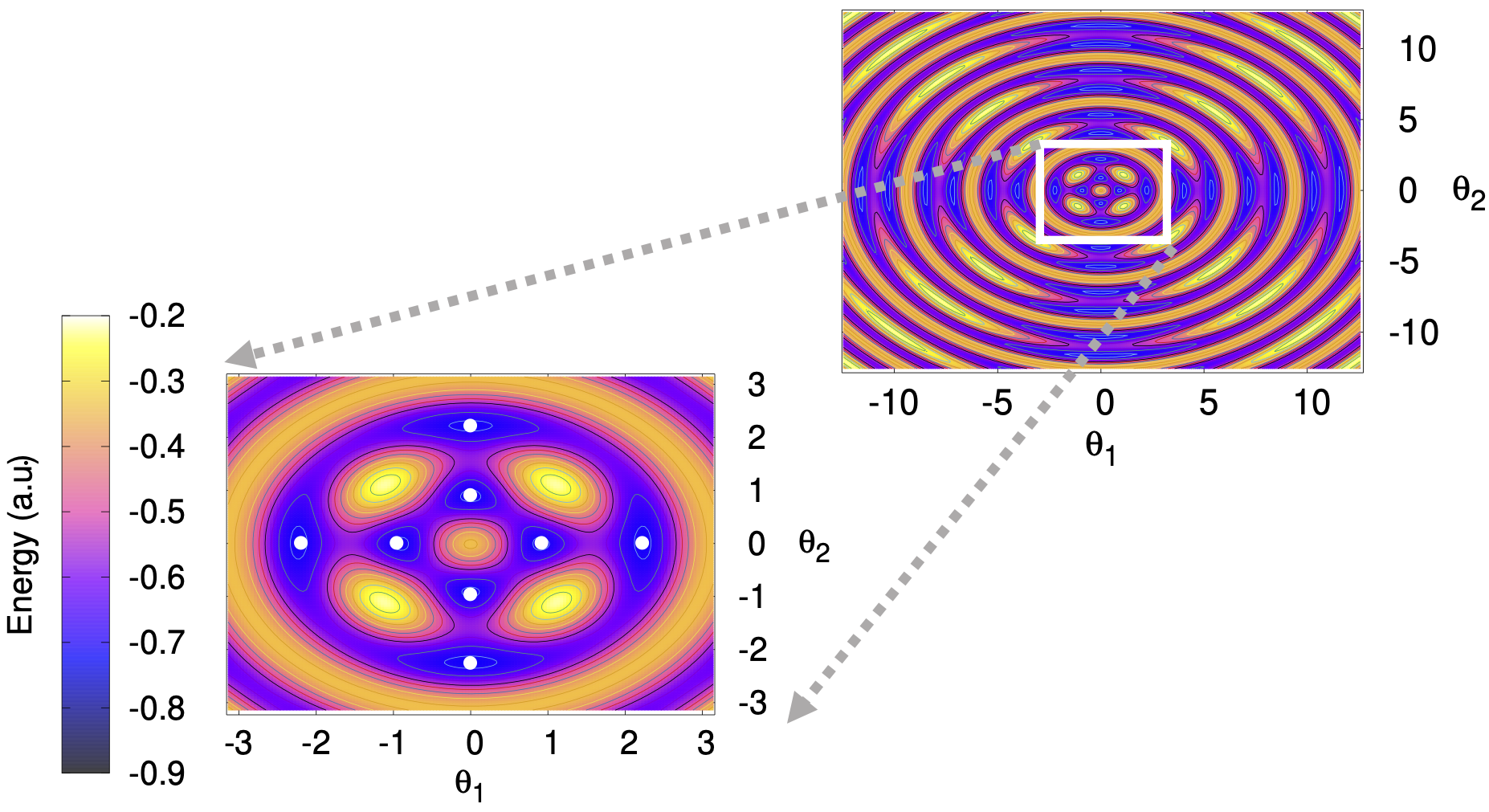}
    \caption{The H$_{3}^{+}$ HES$ (\theta_{1},\theta_{2}$) surface for fixed bond length at r$_{\text{HH}}=4.36$ \AA. Here the surface is expanded about $P_0=P_{core}$.}
    \label{fig:h3_HES_436}
\end{figure}

We continue with our two electron examples with the iso-electronic H$_{3}^{+}$.  Now, instead of a 1D HES, we now have two parameters that mix the one occupied orbital with the two virtual orbitals.   We plot the HES in fig.~\ref{fig:h3_HES_rot} for a linear configuration with hydrogen atoms separated by 2.5 \AA. There are three
minimums for HES$(\theta_{1},\theta_{2}$). In fig. \ref{fig:h3_HES_436} we give the HES of H$_3^+$ at 4.36 \AA~ where there are several minimums with the same globally optimal value.

In both the case of H$_3^+$ and H$_2$, there is a single occupied (spatial) orbital occupied and $m$ virtual orbitals.  For $m=2$, this leads to an $A_{block}$ generator of the form
\begin{equation}
    A_{block}=\begin{bmatrix}
    0 & -\theta_1 & -\theta_2\\
    \theta_1 & 0 & 0\\
    \theta_2 & 0 & 0
    \end{bmatrix}
    \label{eq:A3}
\end{equation}
The eigenvalues of this matrix are $\lambda_\pm=\{0,\pm i\sqrt{\theta_1^2 + \theta_2^2}\}$.  Since the matrix exponential of $A_{block}$ merely exponentiates the eigenvalues, when $\lambda_\pm=i\pi$, the rotation acts trivially on the density matrix.  This underlies periodicity to the plots seen in \ref{fig:h3_HES_rot} and \ref{fig:h3_HES_436}.  

We can explain the periodicity in terms of this invariant by converting to polar coordinates where $\theta_1=R\cos\phi$ and $\theta_2=R\sin\phi$.  Now the nontrivial eigenvalue is $\lambda_\pm=\pm iR$ and we can express the periodicity of the plots as HES$(\theta_1,\theta_2)$=HES$(R,\phi)=$HES$(R+n\pi,\phi)$ with $n$ an integer.

There is a nice generalization of this fact. For a single spatial orbital that is doubly occupied with two electrons and $m$ virtual orbitals, the generalization of \eqref{eq:A3} is
\begin{equation}
     A_{block} = 
    \begin{bmatrix}
        0 & -\theta_{1} & -\theta_{2} & -\theta_{3} & \dots & -\theta_{m} \\
        \theta_{1} &  0 & 0  & 0 & \dots & 0 \\
        \theta_{2} &  0 & 0  & 0 & \dots & 0 \\
        \theta_{3} &  0 & 0  & 0 & \dots & 0 \\
        \vdots & \vdots  & \vdots & \vdots & \ddots & \vdots  \\
        \theta_{m} &  0 & 0  & 0 & \dots & 0
    \end{bmatrix}
\end{equation}
It is straightforward to calculate that the eigenvalues of this matrix are zero except $\lambda_{\pm}=\pm i \sqrt{ \theta_{1}^{2}+\theta_{2}^{2}+\theta_{3}^{2}+ ... + \theta_{m}^{2}}$.  Following the same argument as in the $m=2$ case above, we can say that
\begin{equation}
    \text{HES}(\Theta)=\text{HES}(R,\Phi) = \text{HES}(R+n\pi, \Phi)
\end{equation}
where $R=\sqrt{\theta_1^2+...+\theta_m^2}$.

Therefore, the range of the minimal required search space for each $\theta_j$ is restricted to a hyper-sphere with radius $\pi$ of dimension $m$. But, the default search space was a hyper cube of dimension $m$ with side $2\pi$. Now, the ratio
of minimal required search space with default search goes to zero as $m$ tends to infinity. 
This is a well known consequence of the vanishing ratio of the volume of a hyper-sphere to the volume of the corresponding hyper-cube~\cite{scott2015multivariate,stanton1968multiple}.
 
\subsection{Convergence analysis}

We used the quantum algorithm outlined in ref.~\cite{Google20} for obtaining RHF solutions for four examples.  Depending on initial guess it may converge to local rather global solutions.

\begin{figure}[h!]
    \centering
    \includegraphics[scale=0.4]{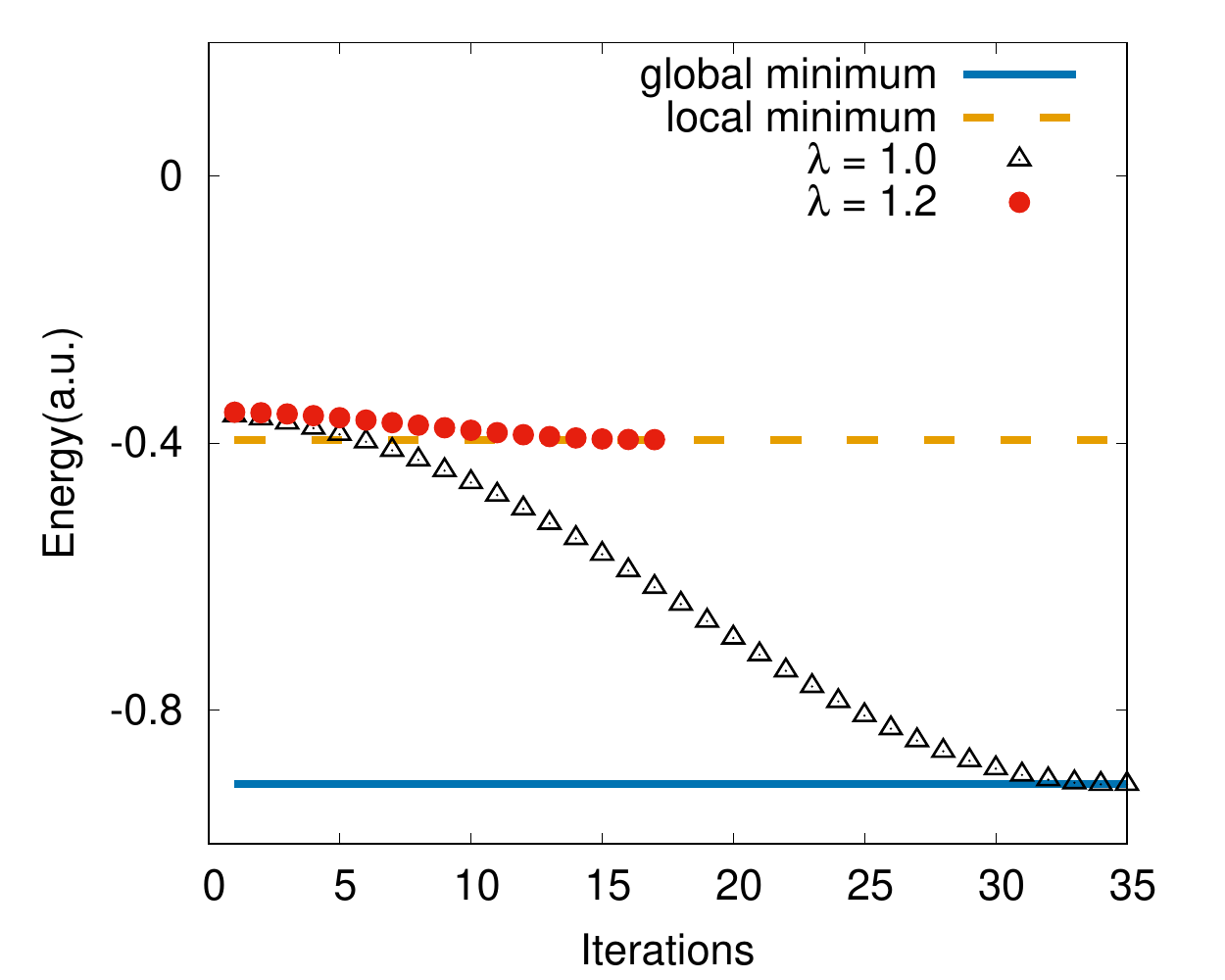}
    \caption{Convergence of the HF quantum optimization to different minimums for the Hartree-Fock instance of H$_{2}$ at r$_{\text{HH}}=1.5$ \AA~ in a minimal basis.  Depicted are two initial conditions obtained by adding a offsets of differing strengths, $\lambda$, to the global ground state. The value used in the Open-Fermion implementation was $\lambda=0.1$.}
    \label{fig:h2_locate_min}
\end{figure}

\begin{figure}[th!]
    \centering
    \includegraphics[scale=0.4]{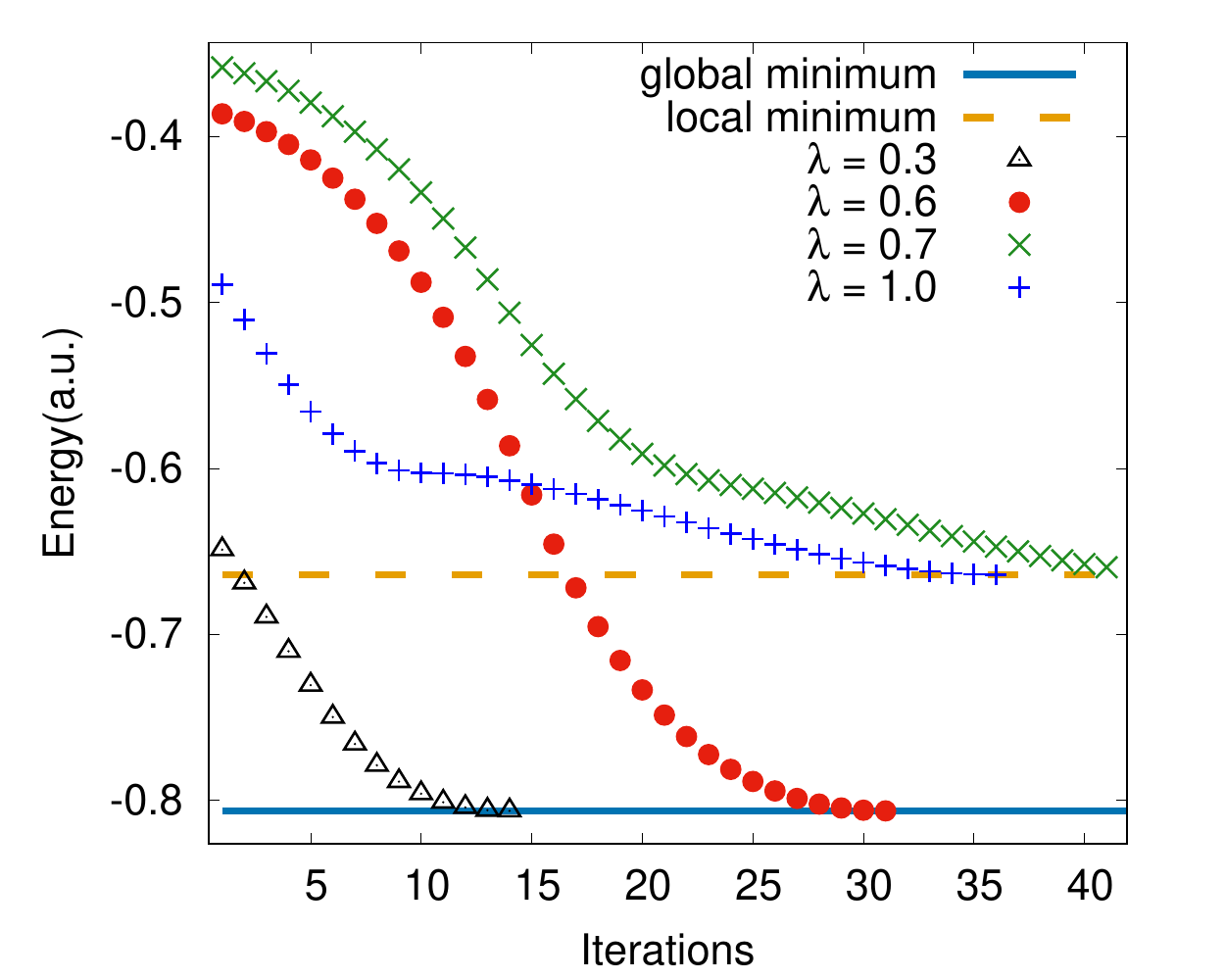}
    \caption{Convergence of the HF quantum optimization to different minimums for the Hartree-Fock instance of H$_{3}^+$ at r$_{\text{HH}}=2.5$ \AA~ in a minimal basis.  Depicted are two initial conditions obtained by adding a offsets of differing strengths, $\lambda$, to the global ground state. }%The value used in the Open-Fermion implementation was $\lambda=0.1$.}
    \label{fig:h3_locate_min}
\end{figure}

The different initial guess were generated using 
the Givens rotations corresponding to different minimums in 
figs.~\ref{fig:h2_HES_rot} and  \ref{fig:h3_HES_rot}, respectively. The results for H$_{2}$, converging to two different minimums on quantum simulator is shown in fig. \ref{fig:h2_locate_min}. 
Similarly results for H$_{3}^{+}$, converging to two different minimums on quantum simulator is shown in fig. \ref{fig:h3_locate_min}. 
In fig. \ref{fig:h2_locate_min}, the values of $\lambda=1$ and $\lambda=1.2$ were chosen as states far enough in parameter space to have energy sufficiently large.  If we select states with small $\lambda$, the convergence to the minimum is highly likely so long as the system does not climb uphill in energy since the initial state has energy less than all minimums except the global minimum.

As final examples, we choose diatomic carbon and its cation.
We also consider C$_2$ and C$_2^{2+}$ as instances that are commonly known to confound solvers due to the appearance of saddle points with in the optimization landscape.  To illustrate the complications of convergence we modified the initial parameters following the method used in \cite{Google20}.  Namely, we begin with the solution provided by the classical SCF solver, perturb from those solutions and observe if the quantum algorithm still converges to the correct minimum.

\begin{figure}[th!]
    \centering
    \includegraphics[scale=0.3]{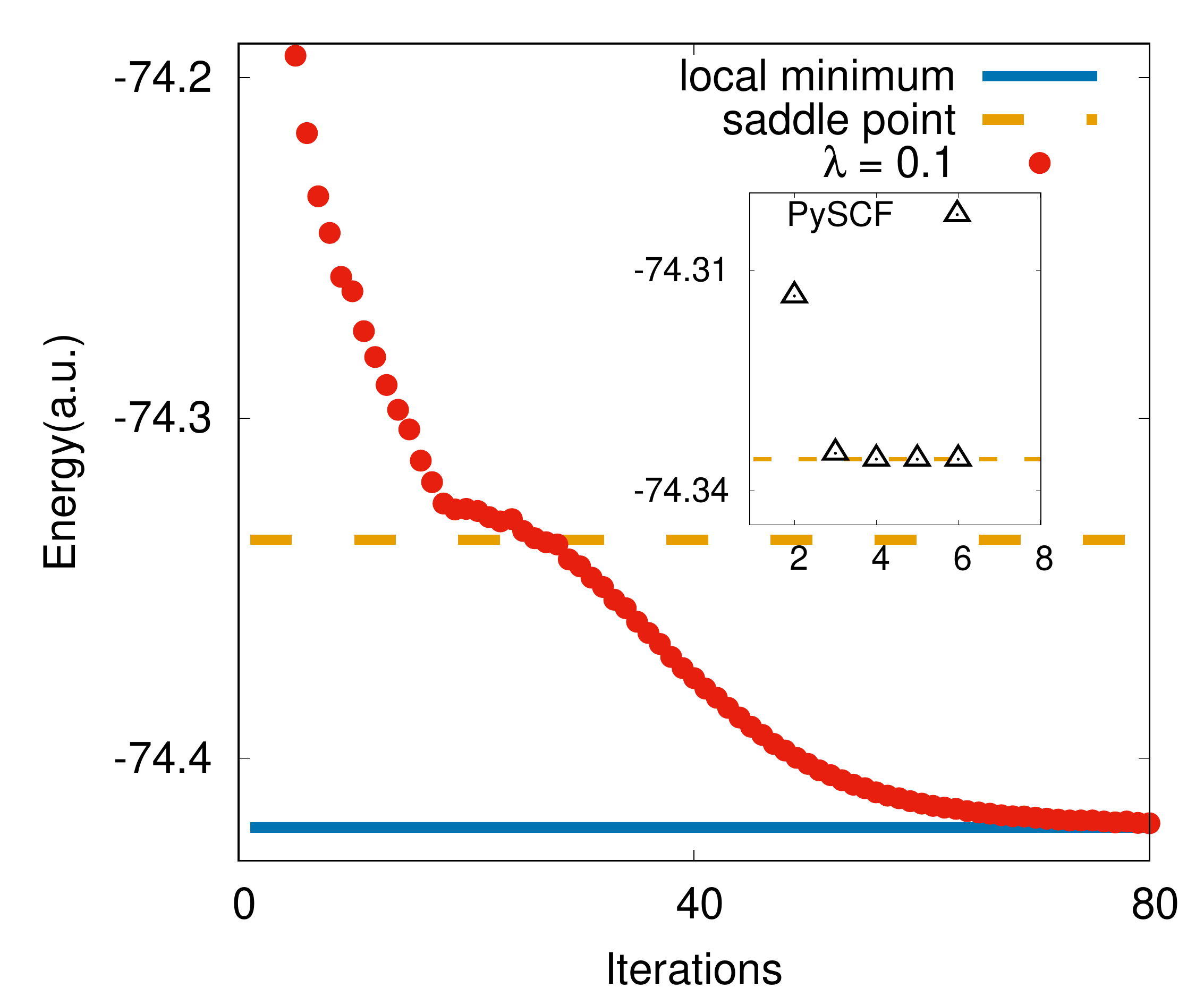}
    \caption{SCF convergence for different optimizers for the minimal basis C$_2$ at r$_{\text{CC}}=1.5$ \AA. Plotted are the optimization trajectories for the quantum circuit optimizer, and PySCF with and without an extra step.  The energy of the saddle point and of the global minimum of the HES are also shown. PySCF guess is $P_{core}$.} 
    \label{fig:c2_qsim}
\end{figure}

\begin{figure}[th!]
    \centering
    \includegraphics[scale=0.3]{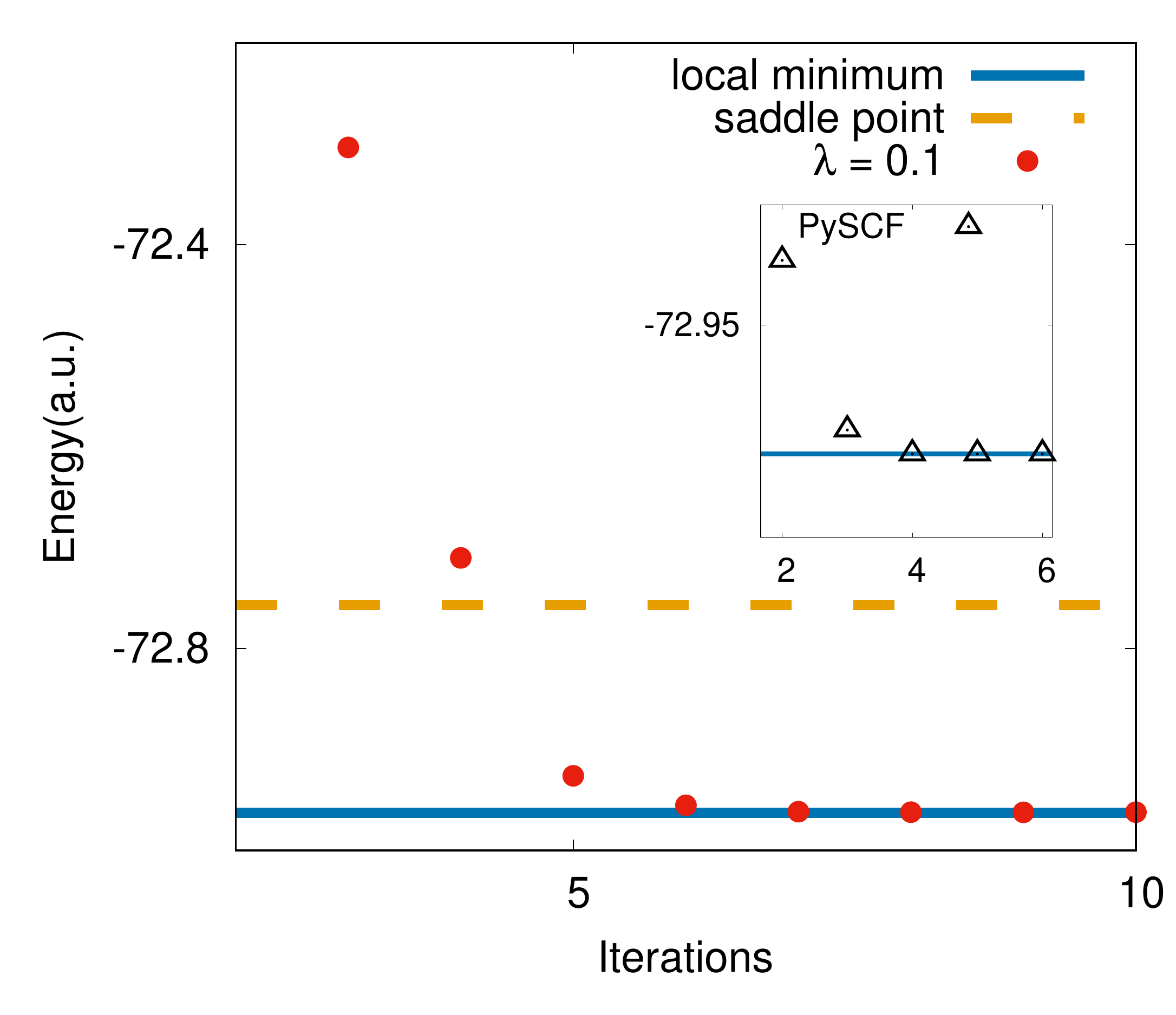}
    \caption{Optimization to local minimums for C$_{2}^{2+}$ on quantum simulator, while avoiding convergence to saddle point at r$_{\text{CC}}=1.0$ \AA. PySCF guess is $P_{core}$.}
    \label{fig:c2_2+_qsim}
\end{figure}

 This example allows us to highlight the importance of using the information of Hessian to avoid saddle points during SCF optimization in either quantum or classical methods.  In figs. \ref{fig:c2_qsim} and \ref{fig:c2_2+_qsim}, convergence results are shown for C$_{2}$ and C$_{2}^{2+}$, respectively.  In each of the plots, we have plotted the performance of the quantum circuit optimization routine and the classical optimization routines as implemented in PySCF.  At an inter-nuclear separation of $1.5$ \AA, there is a saddle point appearing the potential energy landscape.

The use of the orbital Hessian helps solvers avoid saddle points where the gradient may vanish at a non-optimal values.
The solver uses Hessian which 
provides a notion of the curvature of the landscape.  This allows the solver to avoid saddle points.  The augmented Hessian conjugate-gradient method \cite{Sun16} was used for the quantum optimization of the Hartree-Fock circuit.  This allows the solver to avoid convergence to saddle points as illustrated in figs. \ref{fig:c2_qsim} and \ref{fig:c2_2+_qsim}.
Many of the quantum chemistry do not check the  the RHF solutions by default. This makes them prone to failure by convergence to a saddle point rather than the RHF solution.

\section{Discussion}
In this study, we showed that the convergence to a local minimum or global minimum is a function of initial guess. 
The proposed algorithm on quantum simulator carries this feature from classical algorithm. There is no \emph{a priori} guarantee that it will find global solution. 

Since the Hartree-Fock problem is an optimization problem, quantum computing via a modified Grover search \cite{Grover96,Bennett97} can be used to find the global solution quadratically faster than the classical brute force approaches.  However, in both conventional and quantum solvers, local searches are employed for local search whereby no guarantees on finding the globally optimal solutions are given.

In analyzing the Hartree-Fock functional for H$_2$, H$_3^{+}$,
C$_{2}$, and C$_{2}^{2+}$ we note that the number of critical points in the solution space changes as the nuclear separation changes.  This also implies that there are additional difficulties in applying Hartree-Fock for nuclear dynamics or other non-equilibrium configurations.  Most solvers for the Hartree-Fock problem do not explore the entire space and usually only choose a single starting point (rather than multiple starting points).  While this is not often an issue, it can cause serious pathology when used in post-Hartree-Fock methods e.g. coupled cluster method. This will be true for both on conventional computers and in its unitary formulation for quantum computers. 

\section{Conclusions}

The recent application of quantum technology to the Hartree-Fock problem may serve as a hardware benchmark but is unlikely to have a dramatic impact on the practical approaches to this problem.

While the application to Hartree-Fock is not likely to change the workhorse routines used on conventional computers, there are still interesting use cases for the results from Ref.~\cite{Google20}. The projection methods and purity extrapolation presented in \cite{Google20} will still be useful.  

Hartree-Fock is NP-hard and is not likely to admit more than a quadratic speed up.  When considering application areas of quantum computers, it is far more likely to make major breakthroughs when considering time-dependent phenomena.  For example, the quantum-classical hybrid algorithm for obtaining the Kohn-Sham potential of time-dependent density functional \cite{Yang19,brown19,Whitfield14_tddft}, also requires measuring the bond density matrix. 

This article highlights the wide body of knowledge on the SCF method, its difficulty, and shows the lack of verification of the solution are material in both the quantum and conventional computing domains.  

\section{Acknowledgements}
The authors thanks Z. Zimboras and N. Anand for many useful discussions. JDW acknowledges support from Department of Energy Grant DE-SC0019374 "Quantum Chemistry for Quantum Computers."  Additional support for JDW comes from the NSF (PHYS-1820747, EPSCoR-1921199) and from the U.S. Department of Energy, Office of Science, Office of Advanced Scientific Computing Research under programs Quantum Computing Application Teams and Accelerated Research for Quantum Computing program. SG would like to thank N. Anand for useful discussions and USC for the Dornsife Graduate School Fellowship.

\bibliography{hfqc}% Produces the bibliography via BibTeX.

\end{document}